# Crystal Structure, Magnetic Properties and Bonding Analysis of $M_3Pt_{23}Ge_{11}$ (M=Ca, Sr, Ba and Eu)

*Xin Gui and Robert J. Cava*[*]

[1]Department of Chemistry, Princeton University, Princeton NJ 08540, USA

***ABSTRACT***

The properties of Pt-based materials can be intriguing due to the importance of spin-orbit coupling for Pt. Herein, we report four new phases with formulas $M_3Pt_{23}Ge_{11}$ (M = Ca, Sr, Ba and Eu), which adopt the same structure type as $Ce_3Pt_{23}Si_{11}$. Magnetic susceptibility measurements indicate that none of the phases is superconducting above 1.8 K, while for $Eu_3Pt_{23}Ge_{11}$ ferromagnetic ordering is observed at ~ 3 K. The low Curie temperature for that material compared to that of $Eu_3Pt_{23}Si_{11}$ may be due to its larger Eu-Eu distance. One potential factor that destabilizes the structure of other rare-earth based $M_3Pt_{23}Ge_{11}$ is demonstrated through COHP calculations.

*Address correspondence to: rcava@princeton.edu

## 1. Introduction

In some materials, strong spin-orbit coupling plays an essential role in determining the electronic properties,[1–5] making heavy-element-based materials of interest for continuing study.[6–9] Compounds based on Pt, one of the heaviest metals in the periodic table and the focus of this study, therefore deserve broad attention due to their potentially intriguing properties. The low-temperature noncentrosymmetric heavy fermion superconductor, $CePt_3Si$, has been reported, for example, to antiferromagnetically order below its superconducting transition temperature $T_c \sim 0.75$ K.[10] Two other noncentrosymmetric Pt-rich superconductors, $Li_2Pt_3B$ and $LaPt_3Si$, have also been reported.[11,12] And a Pt-rich heavy-fermion superconductor, $UPt_3$, has been reported to display two distinct superconducting phases below 1 K, and an antiferromagnetic ordering transition temperature $T_N \sim 5.8$ K.[13,14]

To broaden the investigation of Pt-based intermetallics, here we report the synthesis, structural characterization and magnetic properties of previously unreported phases $M_3Pt_{23}Ge_{11}$ (M = Ca, Sr, Ba and Eu). Even though the isostructural $RE_3Pt_{23}Si_{11}$ (RE = rare-earth) phases[15] have been reported, considering that in general, the inclusion of magnetic RE might be harmful for superconductivity, alkali-earth (AE)-based platinum germanides were made and characterized to look for superconductivity. Magnetic susceptibility measurements show no evidence for superconductivity above 1.8 K unfortunately. Ferromagnetic ordering is observed at around 3 K for $Eu_3Pt_{23}Ge_{11}$, however. Finally, we compare, through calculations of the total energy, the theoretical structural stabilities of reported $Sm_3Pt_{23}Si_{11}$ and imaginary $Sm_3Pt_{23}Ge_{11}$ phases.

## 2. Experiment

**2.1. Preparation of polycrystalline samples of $M_3Pt_{23}Ge_{11}$:** Starting materials to synthesize $M_3Pt_{23}Ge_{11}$ (M = Ca, Sr, Ba and Eu) were elemental calcium (>99.99%, granules, Beantown Chemical), strontium (99%, granules, Beantown Chemical), barium (>99%, rod, Alfa Aesar), europium (>99.9%, ingot, Beantown Chemical), platinum (99.98%, ~60 mesh, Alfa Aesar) and germanium (99.9999%, pieces, Alfa Aesar). Platinum powder was first pressed into a pellet and arc melted to avoid the mass loss of Pt. The resulting Pt appears as silver chunk. A PtGe precursor was then made *via* arc melting to prevent the mass loss of Ge due to its relatively low vapor pressure considering the experimental condition (>2500 ºC, 50 kPa). Elemental M, PtGe and Pt chunk were then mixed stoichiometrically and placed under a high purity, Zr-gettered, argon atmosphere and arc melted three times, with the melted buttons turned over between melts. All samples were stable in capped glass vials for months. The use of the same procedure for M = Nd, Sm, Dy, Tm and Yb did not result in any indication of the desired product. The samples after arc melting were not annealed since the Ca-, Ba- and Eu-based materials showed good purity with only a very small amount of $Pt_2Ge$ present. Thus they are essentially congruently melting - as-melted materials bring very often used in the research community in materials characterization. $Pt_2Ge$ is not magnetically ordered or superconducting above 1.8 K, so the small amount present in some samples does not affect the property characterization significantly.

**2.2. Phase Identification and Structure Determination:** A Bruker D8 Advance Eco with Cu Kα radiation and a LynxEye-XE detector was employed to determine the phase purity of all samples and the lattice parameters of Ca-, Sr- and Ba-containing materials were determined through powder X-ray diffraction (PXRD). The PXRD patterns were fitted by the Rietveld method in Fullprof using the crystal structure obtained from the analysis of the single crystal data for $Eu_3Pt_{23}Ge_{11}$.[16]

Small crystals (~0.1 × 0.1 × 0.05 to ~0.1 × 0.1 × 0.1 mm$^3$) obtained from an as-cast $Eu_3Pt_{23}Ge_{11}$ sample were employed for structure determination on a Bruker DUO Apex II diffractometer equipped with Mo radiation ($\lambda_{K\alpha}$= 0.71073 Å). Samples were mounted on a Kapton loop and the single crystal measurement was carried out at 300 K with Ω or Φ width of 0.5° and an exposure time of 10 seconds per image. Data acquisition was made *via* the Bruker SMART software with the Lorentz and polarization effect corrections were done by use of the SAINT program. A numerical absorption correction based on crystal-face-indexing was applied through the use of *XPREP*.[17,18] The structure solution was carried out by using direct methods and full-matrix least-squares on F$^2$ with the SHELXTL package.[19]

**2.3. Physical Property Measurements:** Magnetic susceptibility measurements were performed using a Physical Property Measurement System (Quantum Design PPMS) with a vibrating sample magnetometer (VSM) in the temperature range of 1.8 K to 300 K and, when necessary, under various applied magnetic fields. The magnetic susceptibility was defined as M/H where H is the applied magnetic field in Oe and M is the measured magnetization in emu. Tests for superconductivity were performed under a DC magnetic field of 20 Oe.

**2.4. Electronic Structure Calculations:** The Tight-Binding, Linear Muffin-Tin Orbital-Atomic Sphere Approximation (TB-LMTO-ASA) method was employed, using the Stuttgart code, to calculate density of states (DOS) and the Crystal Orbital Hamiltonian Population (-COHP) curves, which identify bonding and antibonding interactions.[20–22] All results were generated using a convergence criterion of 0.05 meV and a mesh of 800 $k$ points.[23] Overlapping Wigner-Seitz (WS) spheres were employed to fill the space where the potential is treated as spherical, with a combined correction for the overlapping part limited to no larger than 19%. Empty spheres were included. The basis set for the calculations included Sm: 6$s$, 6$p$, 5$d$; Pt: 6$s$, 6$p$, 5$p$, 5$d$; Si: 3$s$, 3$p$, 3$d$, 4$s$ and Ge: 4$s$, 4$p$, 4$d$ wavefunctions. The 5$p$ orbital of Pt, 3$d$ orbital of Si and 4$d$ orbital of Ge atoms were automatically downfolded, as determined by the TB-LMTO-ASA program.

## *3. Results and Discussion*

**3.1. Crystal Structure and Phase Determination:** The crystal structures of $M_3Pt_{23}Ge_{11}$ (M = Ca, Sr, Ba & Eu) are equivalent to that of $Ce_3Pt_{23}Si_{11}$[15]. $M_3Pt_{23}Ge_{11}$ crystallizes in a cubic structure with space group *F m-3m* (S. G. 225). The crystallographic data obtained from single crystal XRD are listed in Tables 1 and 2 and Table S1 in the Supporting Information.

To visualize the crystal structure, we take $Eu_3Pt_{23}Ge_{11}$ as the example. As can be seen in Figure 1(a), six edge-shared Eu@Pt$_8$ cuboids form a cage-like structure enclosing one Ge$_4$ tetrahedron and one Pt$_4$ tetrahedron, which are embedded in a larger Ge$_4$ tetrahedron (Figure 1(b). upper). Other than that, Ge and Pt atoms construct two distinct types of polyhedra, a Ge@Pt$_7$ heptahedron and a Ge@Pt$_8$ cube, as shown on the bottom of Figure 1(b). In Figure 1(c), the two types of Ge@Pt polyhedra are seen to build up another cage-like framework with Eu atoms located on the edges and body center of an imaginary cube, as indicated by yellow dash lines.

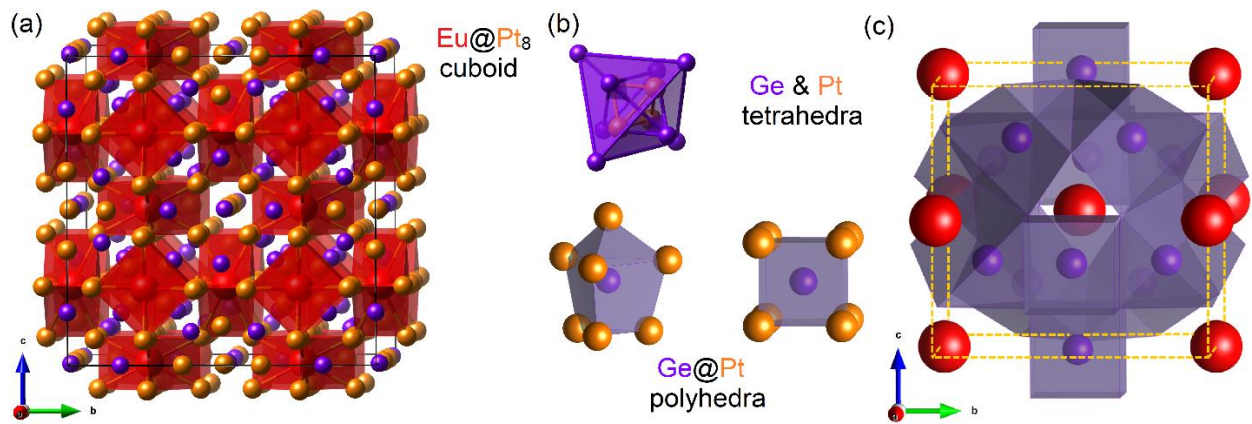

**Figure 1. (a).** The crystal structure of $Eu_3Pt_{23}Ge_{11}$, where red, orange and purple balls represent Eu, Pt, and Ge atoms, respectively. **(b). (top)** $Ge_4$- and $Pt_4$-tetrahedra; **(bottom)** two types of Ge@Pt polyhedra. **(c).** The framework of Ge@Pt polyhedra - the yellow dashed line marks a cube for reference.

The phase purity of the polycrystalline $M_3Pt_{23}Ge_{11}$ materials was determined by powder X-ray diffraction (PXRD), as shown in Figure 2. The patterns were fitted through minor adjustments of the unit cell dimensions obtained from the single crystal diffraction data for $Eu_3Pt_{23}Ge_{11}$. For M = Ca, Sr and Eu, the PXRD patterns showed relatively pure phases, with $Pt_2Ge$ as minor impurity for M = Ca and Eu. However, for $Ba_3Pt_{23}Ge_{11}$, some unidentified peaks appear while the majority of the peaks are fit with the calculated pattern. The relatively impure nature of the Barium variant can be due either to Ba loss during synthesis or due to the fact that the material is not congruently melting. The lattice parameters generated from the refined patterns are shown in Table 3. The lattice parameter of $Eu_3Pt_{23}Ge_{11}$ obtained from refined PXRD pattern is comparable to what is shown in single crystal XRD results.

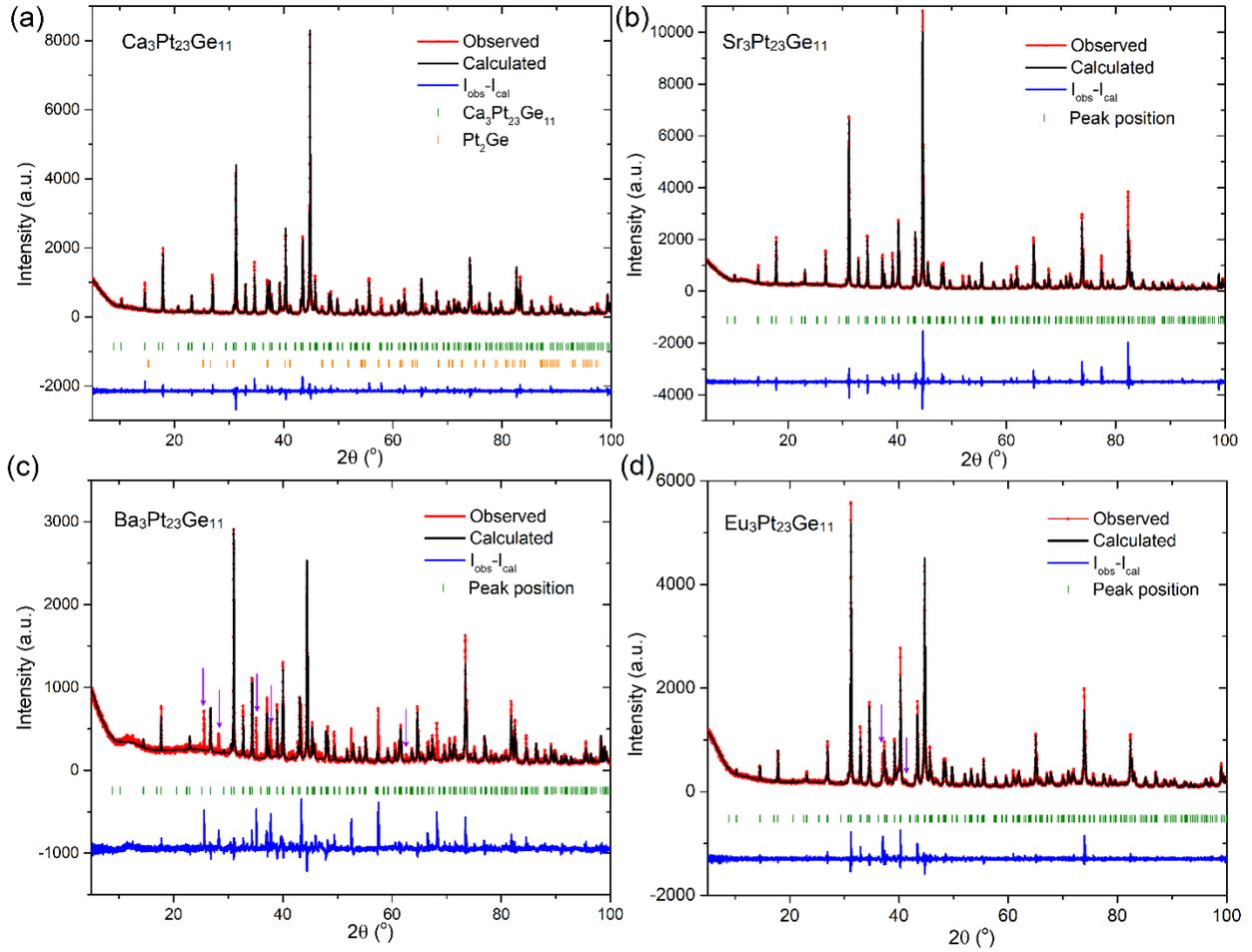

**Figure 2.** Powder XRD patterns with Rietveld fitting of **(a).** $Ca_3Pt_{23}Ge_{11}$; **(b).** $Sr_3Pt_{23}Ge_{11}$; **(c).** $Ba_3Pt_{23}Ge_{11}$; **(d).** $Eu_3Pt_{23}Ge_{11}$.

**Table 1.** Single crystal crystallographic data for $Eu_3Pt_{23}Ge_{11}$ at 293 (2) K.

| Refined Formula | $Eu_3Pt_{23}Ge_{11}$ |
| --- | --- |
| molar density (g/mol) | 5741.44 |
| Space group; Z | $Fm\text{-}3m$; 8 |
| $a$(Å) | 17.1783 (3) |
| V (Å$^3$) | 5069.2 (3) |
| θ range (º) | 2.053-28.265 |
| Extinction coefficient | 0.000029 (1) |
| No. reflections; $R_{int}$ | 10520; 0.0806 |
| No. independent reflections | 372 |
| No. parameters | 29 |
| $R_1$: $\omega R_2$ ($I>2\delta(I)$) | 0.0165; 0.0326 |
| Goodness of fit | 1.224 |
| Diffraction peak and hole (e$^-$/ Å$^3$) | 1.900; -1.433 |

**Table 2.** Atomic coordinates in space group *Fm*-3*m* and equivalent isotropic displacement parameters for $Eu_3Pt_{23}Ge_{11}$. ($U_{eq}$ is defined as one-third of the trace of the orthogonalized $U_{ij}$ tensor ($Å^2$)).

| Atom | Wyckoff. | Occ. | x | y | z | $U_{eq}$ |
|---|---|---|---|---|---|---|
| Pt1 | 32*f* | 1 | 0.08250 (2) | 0.08250 (2) | 0.08250 (2) | 0.0067 (2) |
| Pt2 | 32*f* | 1 | 0.30798 (2) | 0.30798 (2) | 0.30798 (2) | 0.0067 (2) |
| Pt3 | 24*e* | 1 | 0.37371 (4) | 0 | 0 | 0.0083 (2) |
| Pt4 | 96*k* | 1 | 0.08416 (2) | 0.08416 (2) | 0.25335 (2) | 0.0084 (1) |
| Eu5 | 24*d* | 1 | 0 | ¼ | ¼ | 0.0110 (2) |
| Ge6 | 24*e* | 1 | 0.1717 (1) | 0 | 0 | 0.0069 (4) |
| Ge7 | 32*f* | 1 | 0.16468 (6) | 0.16468 (6) | 0.16468 (6) | 0.0069 (4) |
| Ge8 | 32*f* | 1 | 0.39420 (6) | 0.39420 (6) | 0.39420 (6) | 0.0075 (3) |

**Table 3.** Refined face centered cubic lattice parameters for $M_3Pt_{23}Ge_{11}$ (M = Ca, Sr, Ba and Eu).

| M | a (Å) | V (Å³) |
|---|---|---|
| Ca | 17.15018 (8) | 5044.4 (1) |
| Sr | 17.2074 (3) | 5095.0 (3) |
| Ba | 17.2858 (7) | 5165.0 (6) |
| Eu | 17.18691 (3) | 5076.8 (1) |

**3.2. Temperature Dependence of the Magnetic Susceptibility:** The temperature dependence of the magnetic susceptibility for the $M_3Pt_{23}Ge_{11}$ phases was determined from 1.8 to 300 K while general tests for superconductivity using the zero-field cooling method were carried out for M = Ca, Sr and Ba, as shown in Figure S1 in the Supporting Information. No evidence for superconductivity could be found above 1.8 K for any of these materials.

When the applied magnetic field was set as 0.3 T, $Ca_3Pt_{23}Ge_{11}$ showed diamagnetic behavior above ~60 K, below which the magnetic susceptibility turned positive and increased with decreasing temperature, as can be seen in Figure 3(a). For $Ba_3Pt_{23}Ge_{11}$, the signal was always positive while for $Sr_3Pt_{23}Ge_{11}$, the susceptibility sunk to negative values in the temperature range of ~80 - ~150 K before rising again at low temperature. These systematic results indicate that for the materials based on Ca,Sr and Ba, the paramagnetic contribution to the core diamagnetism increases with increasing cell size. The low temperature upturns fitted to the Curie law can be explained by the presence of approximately 1.6, 1.9 or 2 % of weakly interacting spin 1 impurities per formula unit. As shown in Figure 3(d), the magnetic susceptibility of $Eu_3Pt_{23}Ge_{11}$ is always positive from 1.8 K to 280 K, increases rapidly at low temperature and tends to reach a plateau at ~ 2 K, which we interpret as indicating the presence of ferromagnetic ordering. By fitting the inverse magnetic susceptibility curve between 25 K and 150 K with Curie-Weiss law: $\chi = C/(T-\theta_{CW})$, where $\chi$ is the magnetic susceptibility, C is the Curie constant, and $\theta_{CW}$ is the Curie-Weiss temperature, the $\theta_{CW}$ is 3.04 (1) K and the effective moment $\mu_{eff} = \sqrt{8C}$ $\mu_B$ = 14.3 $\mu_B$/f.u.. The inset of Figure 3(d) on the right bottom corner shows the first derivative of the $\chi T$ vs T curve, where the peak shows the ferromagnetic ordering transition temperature (~3.6 K). Compared to the Curie temperature reported for isostructural $Eu_3Pt_{23}Si_{11}$ ($T_C$ = 5.5K)[15], the ferromagnetic $T_C$ is lower in the current case - a result of the replacement of Si by Ge, which increases the shortest Eu-Eu distance and weakens the exchange interaction between Eu atoms.

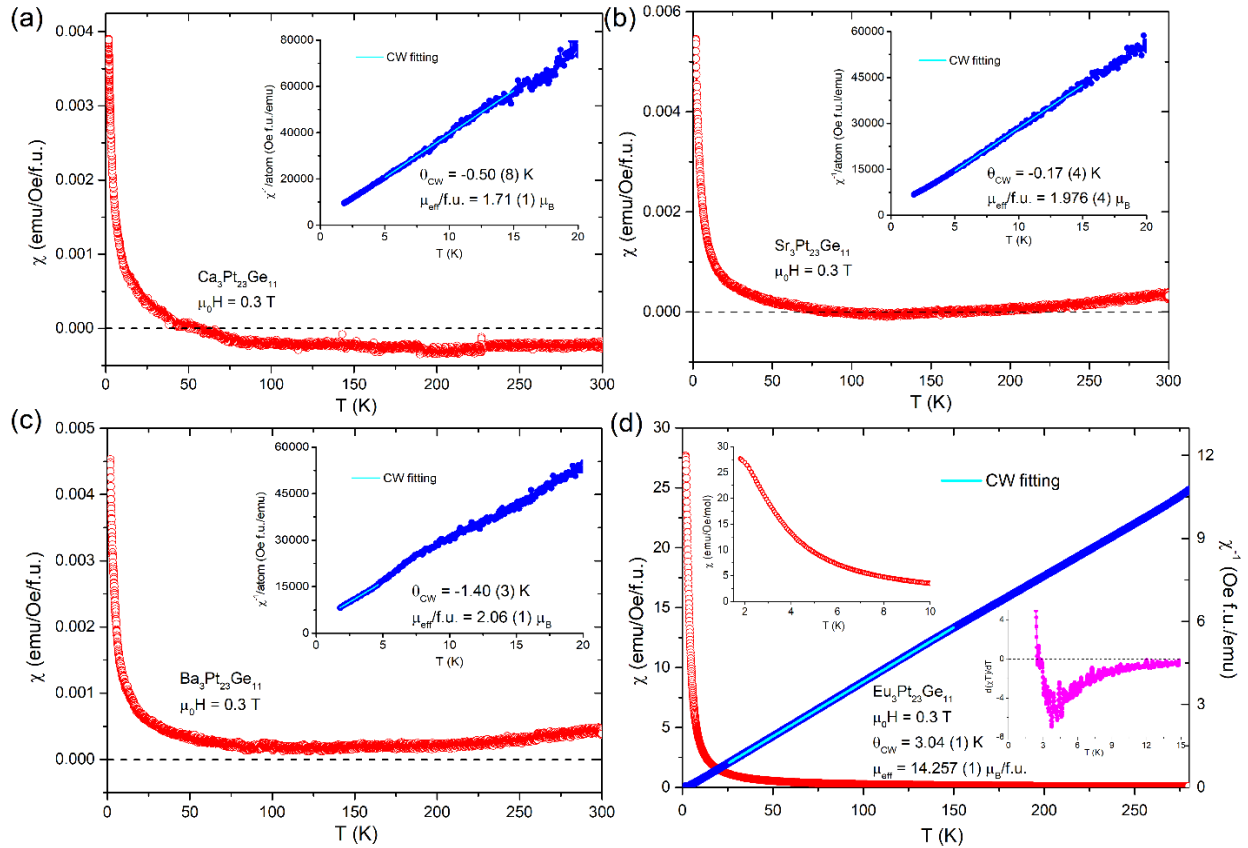

**Figure 3.** Temperature-dependence of the magnetic susceptibility for **(a).** $Ca_3Pt_{23}Ge_{11}$; **(b).** $Sr_3Pt_{23}Ge_{11}$; **(c).** $Ba_3Pt_{23}Ge_{11}$; **(d).** $Eu_3Pt_{23}Ge_{11}$. The insets in **(a)**, **(b)** & **(c)** show Curie-Weiss fitting for low-temperature data (< 20 K). The insets in **(d)** show **(top)** the low-temperature region of the χ vs. T curve and **(bottom)** the first derivative of the χT vs. T curve at low temperatures.

**3.3. Field Dependence of the Magnetization:** The field dependence of the magnetization of the $M_3Pt_{23}Ge_{11}$ materials was measured from 0 to 9 T for M = Ca, Sr and Ba, and -9 to 9 T for M = Eu. As can be seen in Figure 4(a), (b) & (c), small, saturated moments can be observed in the $AE_3Pt_{23}Ge_{11}$ materials (AE = alkaline earth.) This results from the unpaired weakly interacting electrons present, likely from magnetic impurities, as previously described. The hysteresis loops for $Eu_3Pt_{23}Ge_{11}$ are illustrated in Figure 4(d). At temperatures where the magnetization saturates, it is clearly orders of magnitude larger than is seen for the alkaline earth materials. The small coercivity (~50 Oe at 2 K) implies a soft ferromagnetic character for this material. By extrapolating the linear part in the high-field region, the saturated moment per Eu atom was determined to be ~7.0 μ$_B$/Eu, which is consistent with isostructural $Eu_3Pt_{23}Si_{11}$[15], a reflection of the S = 7/2 state for $Eu^{2+}$.

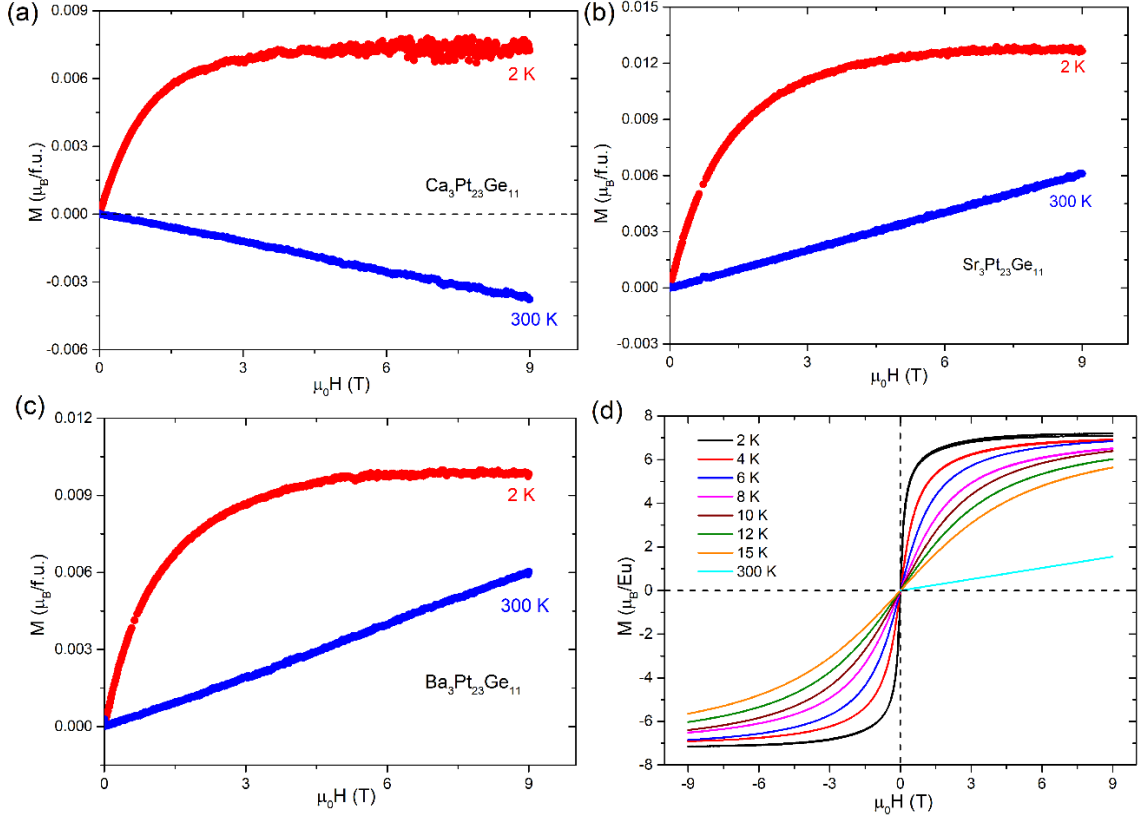

**Figure 4.** Field-dependence of the magnetization for **(a).** $Ca_3Pt_{23}Ge_{11}$; **(b).** $Sr_3Pt_{23}Ge_{11}$; **(c).** $Ba_3Pt_{23}Ge_{11}$; and **(d).** $Eu_3Pt_{23}Ge_{11}$.

**3.4. Analysis of the Structural Stability.** We take the Sm variant as an example due to the difference between Si and Ge formulations. $Sm_3Pt_{23}Si_{11}$ has been previously reported (it is ferromagnetically ordered at $T_C$ = 10.1 K[15]) while the Ge version cannot be made *via* the same method. Thus, we speculate that the change of atomic size on Si/Ge site may be the essential factor that prevents $Sm_3Pt_{23}Ge_{11}$ from being a stable phase under arc-melting. Here we utilize the Tight-Binding, Linear Muffin-Tin Orbital-Atomic Spheres Approximation (TB-LMTO-ASA) method to analyze the bonding/anti-bonding conditions for both $Sm_3Pt_{23}Si_{11}$ and $Sm_3Pt_{23}Ge_{11}$ by calculating the Crystal Orbital Hamiltonian Population (-COHP) curves. By looking at bonding/anti-bonding characters for both $Sm_3Pt_{23}Si_{11}$ and hypothetical $Sm_3Pt_{23}Ge_{11}$, one can intuitively understand how the atomic size change from Si to Ge destabilizes the structure. The lattice parameters for hypothetical $Sm_3Pt_{23}Ge_{11}$ were obtained by using the same scaling found when comparing the values reported for $Eu_3Pt_{23}Si_{11}$ and $Eu_3Pt_{23}Ge_{11}$. Figure 5 plots the COHP curves and density of states (DOS) obtained for both $Sm_3Pt_{23}Si_{11}$ and $Sm_3Pt_{23}Ge_{11}$, where positive COHP values indicate bonding interactions and negative values represent antibonding interactions. The Sm-X (X = Si or Ge) interatomic interactions are too weak to dominate the stability and are not discussed. In the low energy part (< -3 eV), bonding interactions for Pt-Pt, Pt-X and Sm-Pt pairs are dominant, which indicates that these interatomic interactions stabilize the structure. For energies approaching the Fermi level ($E_F$), the Pt-Pt and Sm-Pt interactions show more antibonding character while the Pt-X interactions have bonding character. Near $E_F$, Pt-Pt and Sm-Pt show strong antibonding features in both compounds with two antibonding peaks observed just above the Fermi energy. The Pt-X interaction is the only one that shows bonding character among the four types of interactions near $E_F$ in both compounds, which implies that it can be considered the key to stabilizing the structure near the Fermi level. To better understand how the replacement of Si by Ge causes the synthesis failure (it was not possible to synthesize $Sm_3Pt_{23}Ge_{11}$ in the way used for $Sm_3Pt_{23}Si_{11}$), the integrated COHP (ICOHP), which demonstrates how the electrons from different atom pairs are distributed with energy range, is listed in Table 4. From Table 4, one can tell that by changing from Si to Ge, the ICOHPs of all four interactions decrease. The decrease of ICOHP values indicate that the bonding strengths are weakened by replacing Si with Ge. This could be easily attributed to the larger size of Ge atom. The calculations imply that once the ICOHP drops ~7.2% on going from $Sm_3Pt_{23}Si_{11}$ to $Sm_3Pt_{23}Ge_{11}$, the material could become unstable.

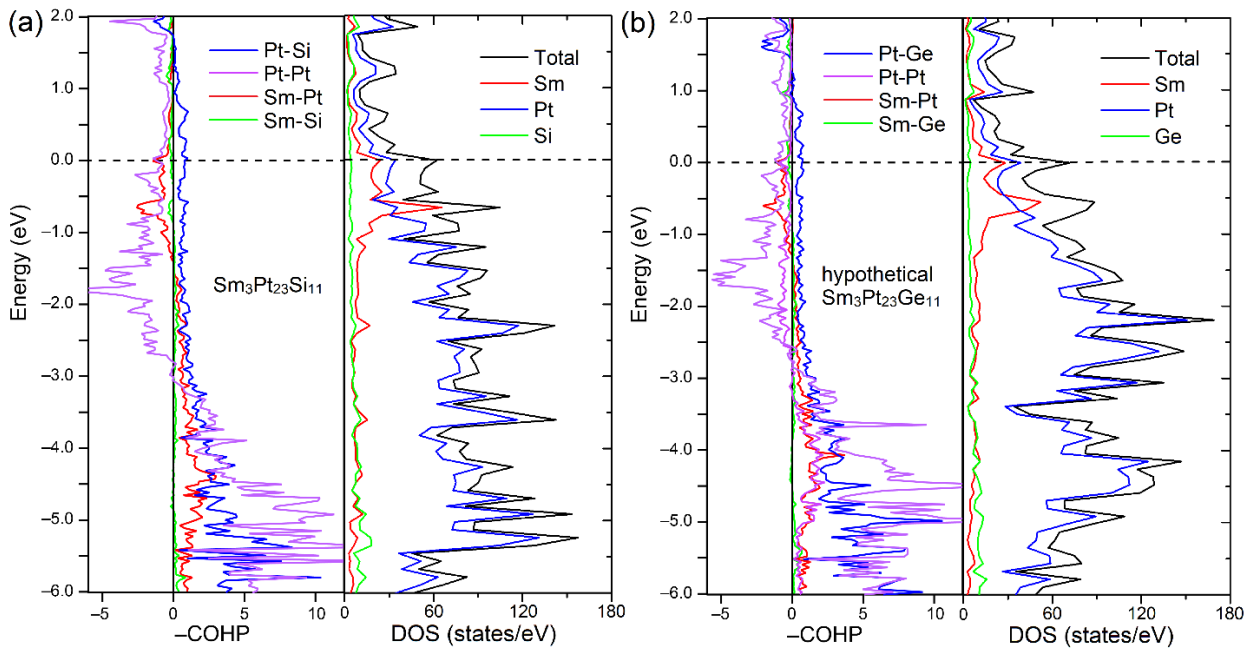

**Figure 5.** Crystal Orbital Hamiltonian Population (-COHP) curves and density of states (DOS) of (a). $Sm_3Pt_{23}Si_{11}$ and (b). hypothetical $Sm_3Pt_{23}Ge_{11}$.

**Table 4.** Integrated COHP (ICOHP) and its percentage (ICOHP %) for $Sm_3Pt_{23}X_{11}$ (X = Si and Ge) below $E_F$.

|  | $Sm_3Pt_{23}Si_{11}$ (ICOHP) | $Sm_3Pt_{23}Ge_{11}$ (ICOHP) |
|---|---|---|
| **Sm-Pt** | 4.363244 | 3.543576 |
| **Sm-X** | 1.268550 | 1.104927 |
| **Pt-Pt** | 23.394574 | 21.521151 |
| **Pt-X** | 32.904933 | 31.300347 |
| **Total** | 61.931301 | 57.470001 |

## *4. Conclusion*

We report here the synthesis of a series of new isostructural compounds $M_3Pt_{23}Ge_{11}$ (M = Ca, Sr, Ba and Eu). Through powder X-ray diffraction and single crystal X-ray diffraction, we found that these new phases are found in the same structure type reported for the Si-based $M_3Pt_{23}Si_{11}$ analogs. The magnetic property measurements showed no observation of superconductivity above 1.8 K for the alkaline-earth-based compounds. $Eu_3Pt_{23}Ge_{11}$ has a Curie Weiss temperature of about 3.0 K and is ferromagnetically ordered at low temperature. Through COHP calculations, we found a plausible explanation for why $Sm_3Pt_{23}Si_{11}$ can be made and $Sm_3Pt_{23}Ge_{11}$ cannot.

## *Acknowledgements*

This work was supported by the US Department of Energy, Division of Basic Energy Sciences, grant DE-FG02-98ER45706.

## *Appendix A. Supplementary Information*

Supplementary information for this article can be found online at xxxxxxx.